# Coupling a Surface Acoustic Wave to an Electron Spin in diamond via a Dark State

D. Andrew Golter[1], Thein Oo[1], Mayra Amezcua[1], Ignas Lekavicius[1], Kevin A. Stewart[2], and Hailin Wang[1]

[1]Department of Physics, University of Oregon, Eugene, Oregon 97403, USA

[2]School of Electrical Engineering and Computer Science,
Oregon State University, Corvallis, OR 97331, USA

## Abstract

The emerging field of quantum acoustics explores interactions between acoustic waves and artificial atoms and their applications in quantum information processing. In this experimental study, we demonstrate the coupling between a surface acoustic wave (SAW) and an electron spin in diamond by taking advantage of the strong strain coupling of the excited states of a nitrogen vacancy center, while avoiding the short lifetime of these states. The SAW-spin coupling takes place through a Λ-type three-level system where two ground spin states couple to a common excited state through a phonon-assisted as well as a direct dipole optical transition. Both coherent population trapping and optically-driven spin transitions have been realized. The coherent population trapping demonstrates the coupling between a SAW and an electron spin coherence through a dark state. The optically-driven spin transitions, which resemble the sideband transitions in a trapped ion system, can enable the quantum control of both spin and mechanical degrees of freedom and potentially a trapped-ion-like solid state system for applications in quantum computing. These results establish an experimental platform for spin-based quantum acoustic, bridging the gap between spintronics and quantum acoustics.

# I. INTRODUCTION

Recent experimental success in coupling surface acoustic waves (SAWs) to a superconducting qubit has led to the emergence of quantum acoustics, the acoustic analog of quantum optics[1]. Acoustic waves propagate at a speed that is five orders of magnitude slower than the speed of light and couple to artificial atoms through mechanical as well as electromagnetic processes, thereby enabling a new paradigm for on-chip quantum operation and communication. The extensive technologies developed for micro-electro-mechanical systems (MEMS) can also be adapted for quantum acoustics. Potential applications in quantum information processing, such as phononic cavity QED, mechanically-mediated spin entanglement, spin squeezing, and SAW-based universal quantum transducers interfacing a broad array of qubits, have been proposed recently[2-6]. Applications for phonon cooling and lasing have also been considered theoretically [7,8]. In addition to superconducting qubits[1,9,10], epitaxially grown as well as gate-defined quantum dots and more recently nitrogen vacancy (NV) centers in diamond have also been coupled to mechanical vibrations including SAWs or mechanical modes of nanomechanical oscillators[11-25].

Among the various artificial atoms explored for quantum acoustics, negatively-charged NV centers in diamond feature exceptional spin properties, including long decoherence times for electron and nuclear spins and high-fidelity optical state preparation and readout[26-32]. The exquisite quantum control of electron and nuclear spin dynamics in NV centers has also led to the realization of quantum state transfer between electron and proximal nuclear spins[33]. The robust spin coherence of NV centers, however, also means that the direct ground-state spin-phonon coupling of a NV center is extremely weak[34,35]. For a nanomechanical oscillator with a mass of several picograms, the ground-state spin-phonon coupling rate at the level of a single-phonon can only reach a few Hz, which is too small for quantum acoustics studies.

In comparison, the orbital degrees of freedom of the excited states of a NV center can couple strongly to long wavelength mechanical vibrations through the induced lattice strain[34-36]. The deformation potential, which characterizes the strength of the electron-phonon coupling, scales with the relevant energy gap and is six orders of magnitude greater than the corresponding parameter for the ground-state spin-phonon coupling. Optomechanical quantum control of a NV center through the excited-state electron-phonon coupling has been demonstrated recently[37]. The strain coupling to the NV excited states has been characterized in detail in a diamond cantilever[38]. Direct excited-state spin-strain coupling has also been investigated[39]. Nevertheless, because of their relatively short lifetime due to optical spontaneous emission, the excited states of NV centers are not suitable for use as qubits.

In this paper, we report experimental studies that use the NV ground states as a spin qubit while taking advantage of the strong excited-state strain coupling to mediate and control the interaction between a spin qubit and a SAW. As illustrated in Fig. 1, the coupling between the spin states and the acoustic wave can take place in a Λ-type three-level system driven by the acoustic wave as well as two external optical fields with a frequency difference of $\omega_B \pm \omega_m$, where $\omega_m$ is the acoustic frequency and $\omega_B$ is the frequency separation between the two spin states. This Λ-type system features a direct dipole optical transition and also a phonon-assisted optical transition, enabled by the excited-state strain coupling. The coupling between an electron spin coherence and a SAW can be controlled via a dark state and probed through phonon-assisted coherent population trapping (CPT). In the limit of sufficiently large detuning for the external optical fields, the upper state in the Λ-type system can be adiabatically eliminated from the dynamics of the two lower states, leading to optically-driven transitions between the phonon ladders of the two spin states (which we refer to as sideband spin transitions), as illustrated in Fig. 1b.

The realization of the phonon-assisted CPT and especially the sideband spin transitions for a NV center establishes an experimental platform for spin-based quantum acoustics. These transitions are analogous to the sideband transitions in the well-known trapped ion system. As shown for the trapped ion system, the sideband spin transitions can enable the quantum control of both the spin and mechanical degrees of freedom, leading to one of the most successful platforms for quantum information processing[40-42]. Combined with diamond nanomechanical oscillators, which can be fabricated in a diamond-on-silicon or bulk diamond approach [43-47], the excited-state mediated and optically-controlled spin-phonon coupling of NV centers opens up a promising avenue to realize a solid-state analog of trapped ion systems.

In the following, we will first discuss the experimental setting, including the generation of SAWs and the formation of the relevant Λ-type three-level systems for a NV center. Next, we will present the experimental demonstration of phonon-assisted CPT. We will then describe the experimental results on sideband spin transitions. The spin transition experiments were carried out in both spectral and time domains. The spectral domain experiment shows that the sideband spin transitions are nuclear spin-selective, enabling the use of the nuclear spins in quantum acoustics. The time domain experiment further probes effects of the upper state excitation on the sideband spin transitions.

## II. EXPERIMENTAL SETTING

For our experimental studies, a NV center situated a few μm below the diamond surface is subjected to incident laser fields and also to a SAW that propagates along and extends approximately one acoustic wavelength below the diamond surface, as illustrated in Fig. 2a. A confocal optical microscopy setup was used for optical excitation and fluorescence collection of a single NV center [48,49]. The experiments were carried out at 8 K, with the diamond sample mounted in a cold-finger optical cryostat. An off-resonant green laser beam ($\lambda$=532 nm) was also used to initialize the NV into the $m_s$=0 ground state.

### A. Generation of surface acoustic waves

SAWs are widely used in electronic devices, such as MEMS devices. A well-established technique for the generation and detection of SAWs is to pattern inter-digital transducers (IDTs) on a piezoelectric substrate. High frequency SAW devices have previously been fabricated on diamond[50]. For the samples used in our experiments, a 400 nm thick layer of ZnO, which is piezoelectric, was first sputtered onto the diamond surface. IDTs were then patterned on the ZnO surface with electron beam lithography. Figure 2b shows an optical image of a pair of IDTs, which are 80 μm apart, patterned on the diamond surface. One IDT, driven by a RF source, serves as a transmitter to generate a SAW via piezoelectric effects. The other, connected to a RF spectrum analyzer, converts the SAW to RF fields and serves as a detector for the characterization of the SAW. Each IDT contains 40 pairs of fingers. The width, $w$, of each finger is 1.5 μm. The center frequency of the SAW generated is given by $v_s/(4w)$, where $v_s$ is the SAW velocity. For our structure, the center frequency measured is near 900 MHz, yielding a SAW velocity of approximately 5600 m/s. The electromechanical coupling efficiency of the IDT is estimated to be approximately 0.05%. An input RF power of 1 W generates a SAW with an amplitude on the order of a pm, as discussed in our earlier work[37].

## B. Λ-type three-level systems in a NV center

In a Λ-type three-level system, an upper state couples to two lower states through two respective dipole optical transitions (see Fig. 1a). As shown in Fig. 3a, a NV center features a ground-state spin triplet, with the $m_s=\pm 1$ states split from the $m_s=0$ state by 2.88 GHz, and six excited states, denoted by $A_{1,2}$, $E_{x,y}$, and $E_{1,2}$ according their respective symmetry properties[34,35]. Dipole optical transitions in a strain free NV center occur between the $m_s=0$ and the $E_{x,y}$ states and also between the $m_s=\pm 1$ and the $A_{1,2}$ and $E_{1,2}$ states. With the $m_s=\pm 1$ states as the two lower states, a Λ-type three-level system can be formed through the $A_2$ transition (states $A_1$ and $E_{1,2}$ can also serve as the upper state, though these states are not as ideal due to various mechanisms of state mixing). In the presence of a DC strain or an external DC electric field, transitions between the $E_y$ and the $m_s=\pm 1$ states, as well as those between the $E_{1,2}$ and the $m_s=0$ states, become allowed due to strain- or electric field-induced state mixing (see Fig. 3b) [34,35]. In this case, a Λ-type three-level system can be formed with the $E_y$ state as the upper state and the $m_s=0$ state and either of the $m_s=\pm 1$ states as the two lower states, as highlighted in Fig. 3b. CPT and optical spin control have been previously observed in Λ-type three-level systems with either the $A_2$ or $E_y$ state as the upper state[48,49,51-53].

For our experimental studies, we have taken advantage of the built-in DC strain in the diamond sample and have used a Λ-type three-level system with the $E_y$ state as the upper state and the $m_s=0$ state as one of the lower states. Excited-state mediated spin-phonon coupling can also be implemented with other Λ-type systems. A technical advantage of using the $m_s=0$ state as one of the lower states is that efficient optical spin detection and initialization can be conveniently implemented without the use of microwave fields.

We have carried out detailed photoluminescence excitation experiments to characterize the optical selection rules and to single out the desired Λ-type three-level system. The inset of Fig. 3c shows the fluorescence collected from a single NV center as a function of the frequency of a probe laser near the zero phonon line ($\lambda \sim 637$ nm). At each data point, the probe laser is periodically (with a period near 10 μs) alternated with a green laser, which reverses ionization and optical pumping. As a result, the population is initialized into the $m_s=0$ ground state. The resulting excitation spectrum features pronounced $E_x$ and $E_y$ resonances. The spitting between the two resonances is 9.6 GHz, indicating a relatively large built-in DC strain[36]. Under this condition, significant strain-induced mixing occurs, which leads to transitions between the $E_y$ and the $m_s=\pm 1$ states as well as transitions between the $E_{1,2}$ and the $m_s=0$ states.

To characterize these processes, we applied a second laser (pump laser), resonant with the $m_s=0$ to $E_x$ transition and measured the fluorescence as a function of the probe frequency. The nearly constant fluorescence level in the excitation spectrum shown in Fig. 3c is due to the pump-driven $E_x$ excitation. This excitation can also pump the electron to the $m_s=\pm 1$ ground states since the $E_x$ state has a small probability of decaying into the $m_s=\pm 1$ states.

The pump-probe excitation spectrum is especially sensitive to the underlying optical pumping process of the probe. As indicated in Fig. 3c, the probe excitation from the $m_s=\pm 1$ ground states leads to three pronounced peaks in the excitation spectrum. These excitations pump the NV back into $m_s=0$ state, increasing the fluorescence from the pump-driven $E_x$ transition. In comparison, the probe excitations from the $m_s=0$ state appear as three dips in the excitation spectrum. These excitations pump the NV from the $m_s=0$ state to the $m_s=\pm 1$ states, effectively reducing the fluorescence from the pump-driven $E_x$ transition. Due to the efficient optical pumping induced by the probe, transitions from the $m_s=0$ to the $E_{1,2}$ states and from the $m_s=\pm 1$ to the $E_y$ states are more pronounced in the pump-probe excitation spectrum shown in Fig. 3c than the probe excitation spectrum shown in the inset of Fig. 3c. This pump-probe excitation spectrum provides detailed information on the optical transitions of the NV center without the usual application of a microwave field.

## III. PHONON-ASSISTED COHERENT POPULATION TRAPPING

As discussed in the introduction, the excited states of NV centers couple strongly to long wavelength lattice strain. This electron-phonon coupling can lead to a strain-induced energy shift and also to state mixing of the relevant excited states [34-36]. For the $E_y$ state, the electron-phonon interaction Hamiltonian, describing the strain-induced energy shift, can be written as:

$$H_{e-phonon} = \hbar g (b+b^+) | E_y \rangle \langle E_y |, \tag{1}$$

where $b$ is the annihilation operator for the phonon mode, $g = D k_m \sqrt{\hbar/2m\omega_m}$ is the effective electron-phonon coupling rate, $D$ is the deformation potential, $k_m$ is the wave number of the phonon mode, and $m$ is the effective mass of the mechanical oscillator. For the phonon-assisted optical transition from the $m_s=0$ to the $E_y$ state and with a laser field at the red sideband of the optical transition (see Fig. 4a), the effective interaction Hamiltonian linear to the mechanical displacement can be derived as (see the Appendix),

$$H_R = \hbar \frac{\Omega_0}{2} \frac{g}{\omega_m} (b | E_y ><m_s|=0| + b^+ | m_s = 0 >< E_y |), \quad (2)$$

where $\Omega_0$ is the Rabi frequency for the laser field coupling to the $m_s=0$ to $E_y$ transition. The effective Rabi frequency for the phonon-assisted optical transition (the red sideband transition) is thus given by $\Omega_R = g\sqrt{n}\Omega_0/\omega_m$, where $n$ is the average phonon number. A similar Hamiltonian can also be derived when the laser is at the blue sideband of the optical transition.

We can use the excited-state electron-phonon coupling to mediate the interactions between spin and mechanical degrees of freedom by incorporating a phonon-assisted optical transition into a Λ-type three-level system, where two ground spin states serve as the two lower states. It is well known that Λ-type three-level systems feature a dark state, a special coherent superposition of the two lower states, which is decoupled from the upper state. This dark state can also be formed for a Λ-type system that incorporates a phonon-assisted optical transition (see the Appendix). Figure 4a shows a Λ-type three-level system, where the upper state, $E_y$, couples to states $m_s=\pm 1$ through a direct dipole optical transition and to state $m_s=0$ through a red sideband transition. For this system, the dark state is given by,

$$|\psi_d> = \frac{1}{\sqrt{\Omega_R^2+\Omega_\pm^2}}(\Omega_R | m_s = \pm 1> - \Omega_\pm | m_s = 0>), \quad (3)$$

where $\Omega_R$ and $\Omega_\pm$ are the Rabi frequencies for the two transitions coupling to the $m_s=0$ and the $m_s=\pm 1$ states, respectively. In this case, the formation of the dark state leads to phonon-assisted CPT, with the electron trapped in the two lower states. This dark state can mediate and control the interactions between the spin states and the relevant phonon mode.

For the phonon-assisted CPT experiment, a SAW with $\omega_m$=818 MHz was coupled to the NV center. The two optical driving fields were derived from a frequency-stabilized tunable dye laser. The detuning between the two laser fields was set by two acoustic optical modulators (AOMs). As indicated in Fig. 4a, one optical field with frequency $\omega_\pm$ drives the direct dipole transition between the $E_y$ and the $m_s=\pm 1$ states. A small magnetic field was also applied to generate a Zeeman splitting, $\omega_B$=24 MHz, between the $m_s=\pm 1$ states. The other optical field with fixed frequency $\omega_0$ was tuned to the red sideband of the $E_y$ transition such that $\omega_0+\omega_m$ is resonant with the $E_y$ transition. This phonon-assisted optical transition was characterized in detail in our earlier work[37].

A NV center, when pumped into the dark state, will remain trapped in the dark state, resulting in the quenching of the NV fluorescence. Figure 4b shows the fluorescence from the $E_y$ state as a function of detuning $\omega_0+\omega_m-\omega_\pm$, for which the NV was initially prepared in the $m_s=0$ state and the powers of the two optical fields were adjusted such that $\Omega_R$ and $\Omega_\pm$ are about equal. Two pronounced dips are observed in Fig. 4b, when the overall Raman resonant condition is satisfied, i.e. when $\omega_0+\omega_m-\omega_\pm$ equals the frequency separation between the $m_s=0$ and the $m_s=\pm 1$ states. Two sets of Λ-type systems are involved in this case. One contains the $m_s=+1$ state and the other the $m_s=-1$ state. The two dips correspond to the quenching of the NV fluorescence when the NV is pumped into the respective dark state. These dips are a direct manifestation of the phonon-assisted CPT process, revealing the coherent interaction between the SAW and the electron spin coherence between the $m_s=0$ and the $m_s=\pm 1$ states.

We have used the density matrix equations for the Λ-type three-level system to model the phonon-assisted CPT experiment (see the Appendix). Figure 4b shows a general agreement between the experiment and the theory. Note that the $m_s=\pm 1$ states exhibit a hyperfine splitting of 2.2 MHz due to coupling with the nitrogen nuclear spin with $I=1$. These hyperfine states, which are included in our model, cannot be clearly resolved in the CPT experiment due to power broadening of the CPT dips. With the assumption that $\Omega_R=\Omega_\pm$, the observed depth of the CPT dip yields $\Omega_R/2\pi=8$ MHz, in agreement with the Rabi frequency estimated from individual dipole optical or phonon-assisted optical transitions[37], and also with the Rabi frequency derived from the sideband spin transition experiment, which will be discussed in the next section.

## IV. OPTICALLY-DRIVEN SIDEBAND SPIN TRANSITIONS

While the phonon-assisted CPT discussed above demonstrates the coherent coupling between a SAW and an electron spin via a dark state, the residual excitation and subsequent decay of the upper state in the Λ-type system introduces an optically-induced decoherence to the spin-phonon system. The upper state excitation can be avoided or reduced through adiabatic passage[54], which has been demonstrated for NV centers[48,55]. For a conventional Λ-type system, the upper state can also be eliminated adiabatically from the dynamics of the two lower states, if the two optical driving fields are sufficiently detuned from the respective dipole optical transitions. In this limit, the three-level system becomes equivalent to an optically-driven spin transition between the two lower states, with an effective Rabi frequency for the spin transition given by $\Omega_S = \Omega_1\Omega_2/(2|\Delta|)$, where $\Omega_1$ and $\Omega_2$ are the Rabi frequencies for the two dipole transitions and $\Delta$ is the dipole detuning, as illustrated in Fig. 1a[48]. In analogy, for a $\Lambda$-type system driven by an acoustic as well as two optical fields and with a sufficiently large detuning for both the optical and the phonon-assisted transitions, the three-level system becomes equivalent to optically-driven spin transitions between the phonon ladders of the two lower states, with an effective Rabi frequency for the sideband spin transition given by

$$\Omega_{SS} = \frac{\Omega_R \Omega_\pm}{2|\Delta|} = \frac{\Omega_0 \Omega_\pm}{2|\Delta|\omega_m} g\sqrt{n} = g_{ss}\sqrt{n}, \qquad (4)$$

as illustrated in Fig. 1b. Here, $g_{ss}$ is the single-phonon Rabi frequency for the sideband spin transition.

We have carried out both spectral and time domain experiments to demonstrate and characterize the sideband spin transitions. For these experiments, optical driving fields similar to those used for the phonon-assisted CPT experiment were used except that we set $\Delta=100$ MHz. A relatively small detuning was employed such that effects of the upper state excitation can still be investigated. The optical and RF powers used for Figs. 5 and 6 are the same as those used for Fig. 4. To detect the spin population in the $m_s=\pm1$ states, we used a resonant laser field to excite the NV from the $m_s=\pm1$ states to the $A_2$ state and measure the corresponding fluorescence, as illustrated in Fig. 5a.

## A. Spectral domain experiment

The pulse sequence used for the spectral domain experiment is shown in Fig. 5b. After the initialization of the NV center into the $m_s=0$ ground state with a 532 nm laser pulse, two optical driving fields with a duration of 2 µs were applied to the $\Lambda$-type three-level system. The SAW field was kept on continuously, since it is far from the relevant resonances and contributes only when the corresponding optical driving field is also on. The population in the $m_s=\pm1$ states was detected via the $A_2$ transition right after the optical driving pulses.

Figure 5c plots the fluorescence from state $A_2$ as a function of detuning $\omega_0+\omega_m-\omega_\pm$, with both $\omega_0$ and $\omega_m$ fixed. This spectrum measures directly the optically-driven, phonon-assisted transitions from the $m_s=0$ to the $m_s=\pm1$ states. As shown in Fig. 5c, the sideband spin transitions take place when the Raman resonant condition is satisfied. The spectral linewidth (full width at half maximum) of the transition resonances is 0.7 MHz, in agreement with the expected spin dephasing rate. The clearly-resolved hyperfine structure of the $m_s=\pm 1$ states, with a 2.2 MHz hyperfine splitting, demonstrates that the sideband spin transitions are nuclear spin selective, thus allowing the use of nuclear spins in quantum acoustics.

**B. Time domain experiment**

To determine the effective Rabi frequency for the sideband spin transitions and also to probe effects of the upper state excitation, we have carried out transient experiments, in which we measure the spin population in the $m_s=\pm 1$ states as a function of the duration of the optical driving fields. As shown in Fig. 6a, the pulse sequence used is the same as that used for the spectral domain experiment except that the duration of the optical driving fields is now a variable parameter.

The solid circles plotted in Fig. 6b show the fluorescence from state $A_2$ as a function of the duration of the optical driving fields, where we fixed $\omega_0$, $\omega_\pm$, and $\omega_m$ such that the Raman resonant condition for the phonon-assisted spin transition was satisfied for the $m_s=+1$ state and for the nitrogen nuclear spin projection $m_n=+1$. The initial rise of the fluorescence is primarily due to the sideband spin transition from the $m_s=0$ state to the $m_s=+1$ and $m_n=0$ state, with the rise time determined by $\Omega_{ss}$. Optical pumping resulting from the excitation and subsequent decay of the upper state can also lead to population in the $m_s=+1$ state. To single out this process, we carried out an experiment with the same experimental condition as that used for the solid circles in Fig. 6b, except that $\omega_0+\omega_m-\omega_\pm$ is 6.5 MHz away from the Raman resonant condition. The experimental result is plotted as the solid squares in Fig. 6b.

The solid lines in Fig. 6b show the theoretical calculations based on the density matrix equations, where we have taken $\Omega_R=\Omega_\pm$ and $\Gamma_1$, the decay rate from the $E_y$ state to the $m_s=\pm 1$ states, as adjustable parameters. The calculations yield $\Gamma_1/2\pi=1.8$ MHz and also $\Omega_R/2\pi=8$ MHz, which agrees with the Rabi frequencies derived from the phonon-assisted CPT experiment using the same optical powers. From these results, we obtain an effective Rabi frequency for the sideband spin transition, $\Omega_{ss}/2\pi=0.3$ MHz. The good agreement between the experiment and theory shows that both the sideband spin transitions and the optical pumping are well characterized by the density matrix equations.

In the limit of large dipole detuning, the upper state population scales with $1/\Delta^2$. Both the optical pumping rate and the optically-induced decoherence rate thus also scale with $1/\Delta^2$. In comparison, $\Omega_{ss}$ scales with $1/\Delta$. In this regard, strong excited-state mediated spin-phonon coupling can be achieved with negligible optical pumping or optically-induced decoherence from the upper state excitation. For example, by setting $\Omega_{\pm}/|\Delta|$ to 1/60, we can keep the optically-induced decoherence rate to about 1 kHz and achieve a single-phonon Rabi frequency, $g_{ss}$, three orders of magnitude greater than that can be achieved with direct ground-state spin-phonon coupling.

## V. SUMMARY AND OUTLOOK

In summary, by coupling a SAW to an electron spin in diamond through a $\Lambda$-type three-level system, we have realized both phonon-assisted CPT and optically-driven sideband spin transitions. These experiments demonstrate that we can take advantage of the strong excited-state electron-phonon interaction and use the $\Lambda$-type three-level system to mediate and control the coupling between spin and mechanical degrees of freedom, thereby establishing an experimental platform for exploring spin-based quantum acoustics. Our approach can also be extended to other emerging material systems with spin defect centers such as SiC[56]. Furthermore, a diamond nanomechanical oscillator featuring optically-driven sideband spin transitions resembles a trapped-ion system, providing a promising system to pursue the highly successful paradigm of trapped-ion based quantum computing in a solid-state system.

## ACKNOWLEDGEMENTS

This work is supported by NSF under grant No. 1414462 and by AFOSR.

# APPENDIX

We consider a Λ-type three-level system, driven by two optical fields and an acoustic field, as shown in Fig. 1b. The two dipole optical transitions, with frequency $\nu_1$ and $\nu_2$, couple to the two optical fields, with frequency $\omega_1$ and $\omega_2$ and Rabi frequency $\Omega_1$ and $\Omega_2$, respectively. With the rotating wave approximation, the Hamiltonian of the system, is given by

$$H = \hbar\omega_m b^+ b - \hbar\nu_1 |g_1\rangle\langle g_1| - \hbar\nu_2 |g_2\rangle\langle g_2| + \hbar g(b^+ + b)|e\rangle\langle e| + \\ \hbar\frac{\Omega_1}{2}(e^{-i\omega_1 t}|e\rangle\langle g_1| + h.c.) + \hbar\frac{\Omega_2}{2}(e^{-i\omega_2 t}|e\rangle\langle g_2| + h.c.) \quad (A1)$$

where $b^+$ and $b$ are the creation and annihilation operators for the acoustic field with frequency $\omega_m$ and $g$ is the electron-phonon coupling rate. Applying the Schrieffer-Wolff transformation

$$U = \exp[\frac{g}{\omega_m}(b^+ - b)|e\rangle\langle e|] \quad (A2)$$

to the Hamiltonian gives

$$\tilde{H} = \hbar\omega_m b^+ b - \hbar\nu_1 |g_1\rangle\langle g_1| - \hbar\nu_2 |g_2\rangle\langle g_2| + \hbar\frac{g^2}{\omega_m}|e\rangle\langle e| + \\ \hbar\frac{\Omega_1}{2}(e^{-i\omega_1 t + \frac{g}{\omega_m}(b^+ - b)}|e\rangle\langle g_1| + h.c.) + \hbar\frac{\Omega_2}{2}(e^{-i\omega_2 t + \frac{g}{\omega_m}(b^+ - b)}|e\rangle\langle g_2| + h.c.) \quad (A3)$$

which has the same form as the trapped ion Hamiltonian[40,57]. Transforming to an interaction picture, we then have

$$\tilde{H}_I = \hbar\frac{\Omega_1}{2}(e^{i\Delta_1 t} e^{-\frac{g}{\omega_m}(b^+ e^{i\omega_m t} - b e^{-i\omega_m t})}|e\rangle\langle g_1| + h.c.) + \\ \hbar\frac{\Omega_2}{2}(e^{i\Delta_2 t} e^{-\frac{g}{\omega_m}(b^+ e^{i\omega_m t} - b e^{-i\omega_m t})}|e\rangle\langle g_2| + h.c.) \quad (A4)$$

where $\Delta_1 = (\nu_1 - g^2/\omega_m) - \omega_1$ and $\Delta_2 = (\nu_2 - g^2/\omega_m) - \omega_2$ are the effective detunings of the two optical fields from their respective dipole transitions.

We assume that the $\Omega_1$ field is tuned near the red phonon-sideband of the $|g_1\rangle$ to $|e\rangle$ transition ($\Delta_1 \approx \omega_m$) and the $\Omega_2$ field is tuned near the $|g_2\rangle$ to $|e\rangle$ transition ($\Delta_2 \approx 0$). Expanding $\tilde{H}_I$ in $g/\omega_m$, which can be viewed as an effective Lamb-Dicke parameter for our solid state system[40], and keeping only the nearly resonant terms, we can approximate the interaction Hamiltonian as

$$H_I = \hbar \frac{\Omega_1}{2} \frac{g}{\omega_m} (be^{i(\Delta_1 - \omega_m)t} |e\rangle\langle g_1| + h.c.) + \hbar \frac{\Omega_2}{2} (e^{i\Delta_2 t} |e\rangle\langle g_2| + h.c.) \quad (A5)$$

which is similar to a Hamiltonian for a Λ-type three-level system driven by two optical fields, with effective detuning, $\Delta_R = \Delta_1 - \omega_m$ and $\Delta_2$, and effective Rabi frequency, $\Omega_R = g\sqrt{n}\Omega_1/\omega_m$ and $\Omega_2$, respectively, where $n$ denotes the average phonon number. Note that the Raman resonant condition is $\Delta_R = \Delta_2$. The above Hamiltonian is also valid for relatively large $\Delta_R$ and $\Delta_2$, as long as $\Delta_R \approx \Delta_2$.

With $\Delta_R = \Delta_2$, the Hamiltonian given by Eq. A5 features a dark state,

$$|\psi_d\rangle = \frac{1}{\sqrt{\Omega_R^2 + \Omega_2^2}} (\Omega_R |g_2\rangle - \Omega_2 |g_1\rangle), \quad (A6)$$

with $H_I |\psi_d\rangle = 0$. This dark state is decoupled from state $|e\rangle$, leading to phonon-assisted CPT of the electron in the two lower state.

The equations of motion for the density matrix elements, $\rho_{ij}$, in the rotating frame, which we have used to model the experiments, can also be derived from $H_I$ and are given as

$$\dot{\rho}_{e1} = -(i\Delta_R + \gamma)\rho_{e1} + \frac{i\Omega_R}{2}(\rho_{ee} - \rho_{11}) - \frac{i\Omega_2}{2}\rho_{21} \quad (A7a)$$

$$\dot{\rho}_{e2} = -(i\Delta_2 + \gamma)\rho_{e2} + \frac{i\Omega_2}{2}(\rho_{ee} - \rho_{22}) - \frac{i\Omega_R}{2}\rho_{12} \quad (A7b)$$

$$\dot{\rho}_{21} = -[i(\Delta_R - \Delta_2) + \gamma_s]\rho_{21} + \frac{i\Omega_R}{2}\rho_{2e} - \frac{i\Omega_2}{2}\rho_{e1} \quad (A7c)$$

$$\dot{\rho}_{ee} = -\Gamma\rho_{ee} + (\frac{i\Omega_R}{2}\rho_{e1} + c.c.) + (\frac{i\Omega_2}{2}\rho_{e2} + c.c.) \quad (A7d)$$

$$\dot{\rho}_{11} = \Gamma_1\rho_{ee} - (\frac{i\Omega_R}{2}\rho_{e1} + c.c.) \quad (A7e)$$

$$\dot{\rho}_{22} = \Gamma_2\rho_{ee} - (\frac{i\Omega_2}{2}\rho_{e2} + c.c.) \quad (A7f)$$

where $\gamma_s$ and $\gamma$ are the decay rates for the spin coherence and optical dipole coherence, respectively, $\Gamma = \Gamma_1 + \Gamma_2$ is the total decay rate for the upper state population, with $\Gamma_1$ and $\Gamma_2$ being the decay rate to $|g_1\rangle$ and $|g_2\rangle$, respectively. For the theoretical calculation, we have used $\gamma_s/2\pi$=0.35 MHz, which is primarily due to spin dephasing induced by the nuclear spin bath, and $\Gamma/2\pi$=14 MHz, as determined experimentally. We have assumed $\gamma=\Gamma/2+\gamma_{orb}$, where $\gamma_{orb}$ is the dephasing rate due to coupling to the orbital degrees of freedom and have taken $\gamma_{orb}$=12 MHz[55]. We have also used $\Gamma_1$=1.8 MHz, as derived from the optical pumping experiment in Fig. 6b. NV spectral diffusion is treated as a Gaussian distribution of the optical transition frequency with a linewidth of 140 MHz, as derived from the excitation spectrum of the $E_y$ resonance.

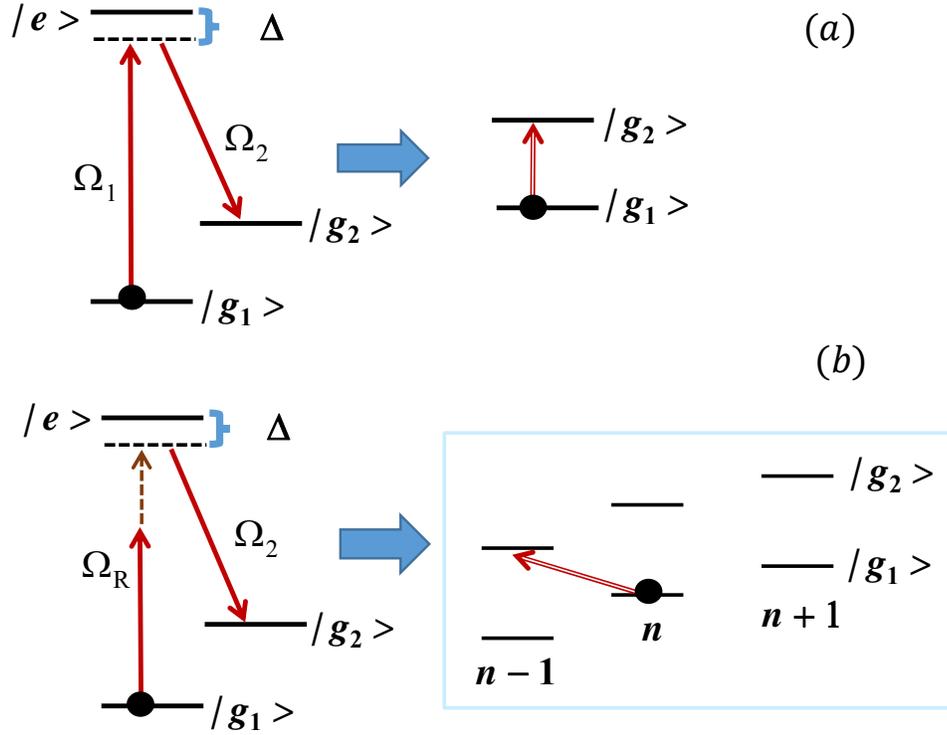

Fig. 1 (a) Schematic of a Λ-type three-level system driven by two optical fields with respective Rabi frequency, $\Omega_1$ and $\Omega_2$. In the limit of large dipole detuning, $\Delta$, the system becomes equivalent to an optically-driven transition between the two lower states. (b) Schematic of a Λ-type three-level system driven by an acoustic field (brown dashed line) as well as two optical fields. The $|g_1\rangle$ to $|e\rangle$ transition is a phonon-assisted transition with effective Rabi frequency $\Omega_R$. In the limit of large $\Delta$, the system becomes equivalent to an optically-driven transition between the phonon ladders of the two lower states, where $n$ denotes the phonon number.

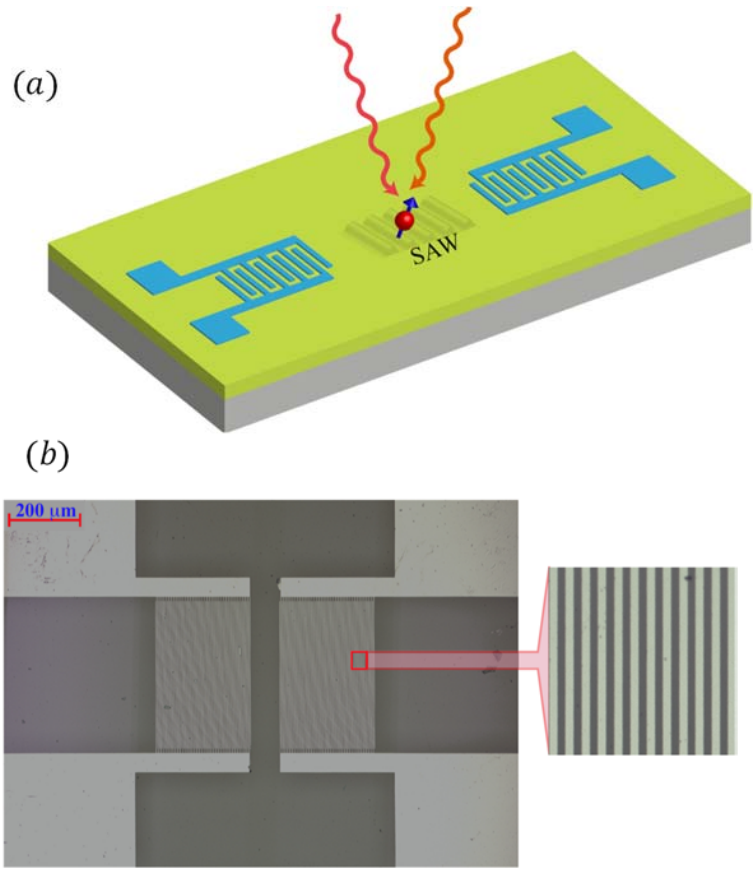

Fig. 2 (a) Schematic of a NV center driven by two optical fields and a SAW field generated by an IDT. (b) Optical image of a pair of IDTs fabricated on the diamond surface.

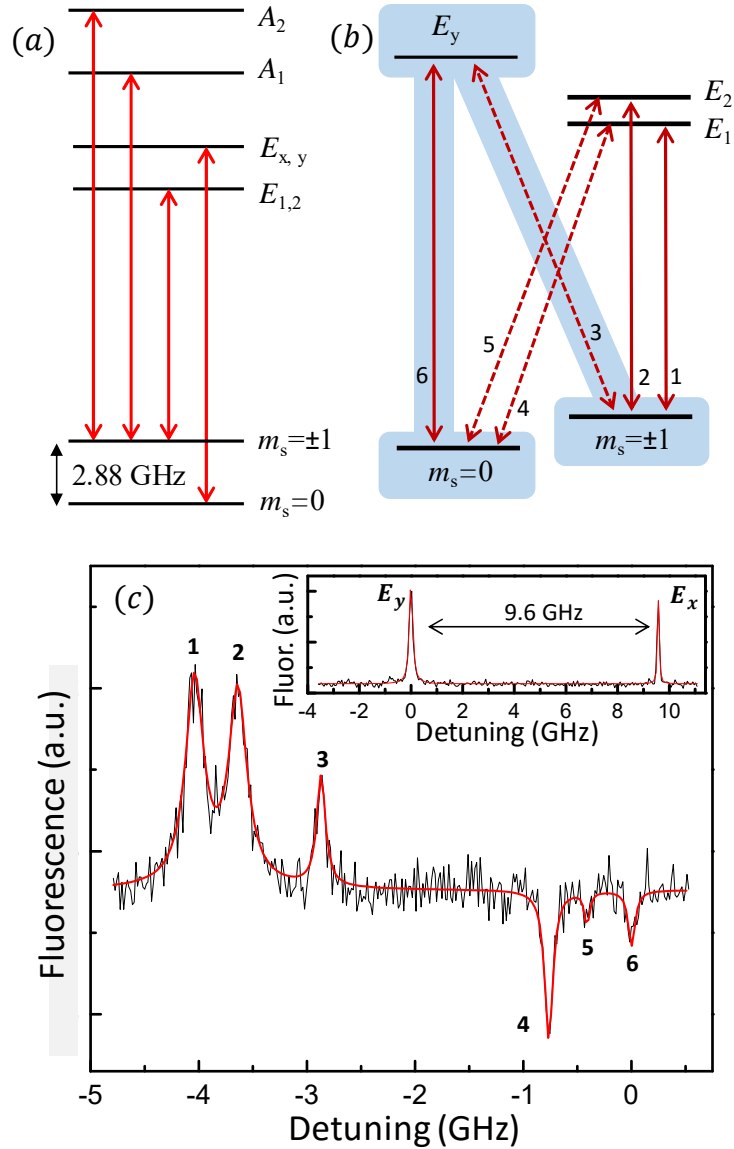

Fig. 3 (a) Energy level structure and dipole optical transitions of a NV center (strain free). (b) State mixing due to built-in DC strain leads to additional dipole transitions, as indicated by the dashed lines. Shaded levels and arrows highlight the Λ-type three-level systems used in our experiments. (c) Excitation spectrum with a pump field fixed at the $E_x$ transition. Peak labels indicate the corresponding transitions in (b). Red lines are fits to Lorentzians. Inset: Excitation spectrum obtained with no pump field.

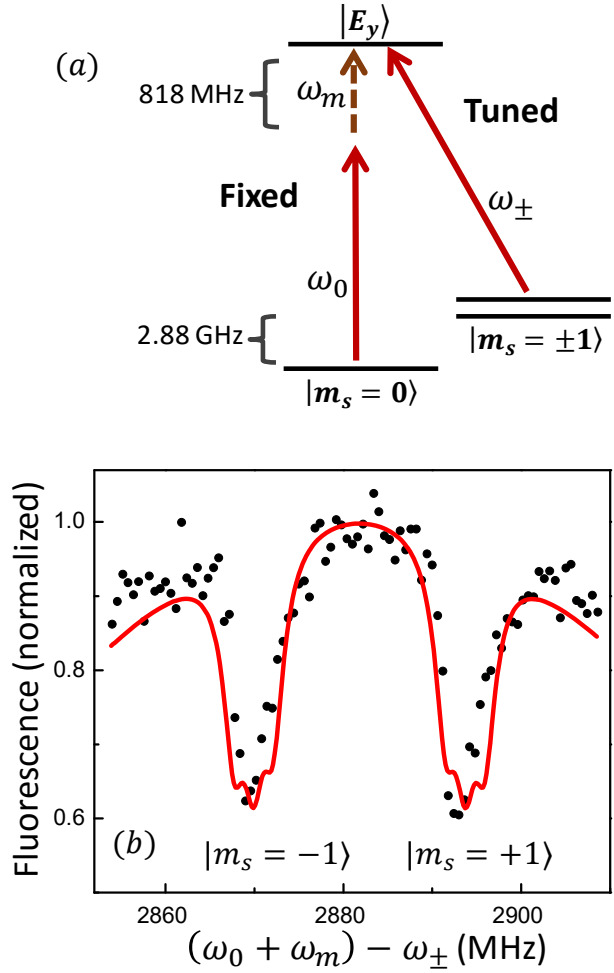

Fig. 4 (a) Energy level diagram used for phonon-assisted CPT. Solid red arrows are the optical fields. Dashed brown arrow is the acoustic field. (b) Fluorescence from state $E_y$ as a function of $\omega_0+\omega_m-\omega_\pm$. The optical power at frequency $\omega_0$ and $\omega_\pm$ is 4 µW and 1 µW, respectively. The RF input power for the IDT is 0.13 W. The two dips correspond to Λ-type systems formed with either the $m_s=+1$ or the $m_s=-1$ state. Solid red curve is the theoretical calculation discussed in the text.

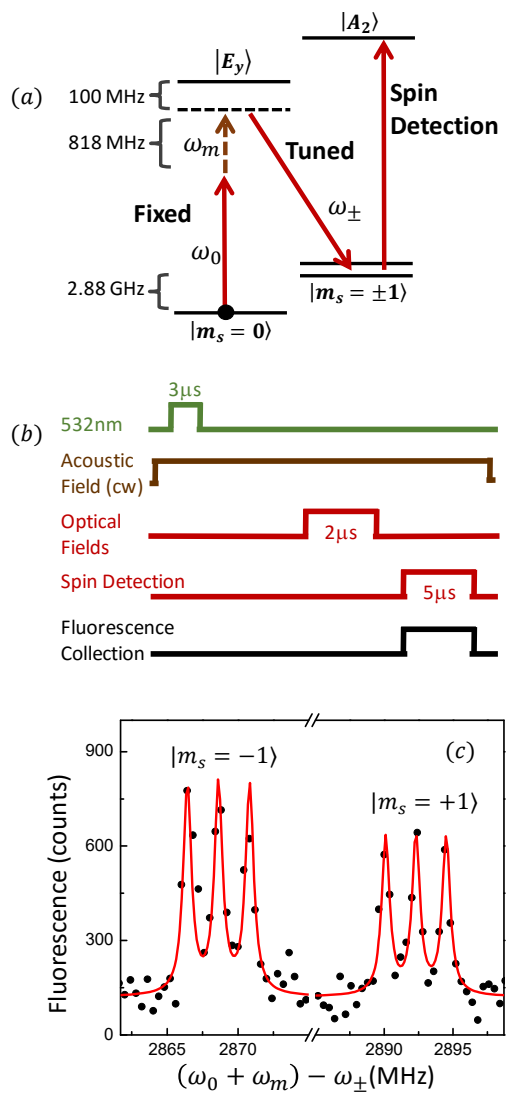

Fig. 5 (a) Energy level diagram including spin detection used for the sideband spin transition experiments. (b) Pulse sequence used for the spectral domain experiment. (c) Fluorescence from state $A_2$ as a function of $\omega_0+\omega_m-\omega_\pm$. Solid red line is a fit to six Lorentzians with equal linewidths. A background due to optical pumping has been subtracted from the data.

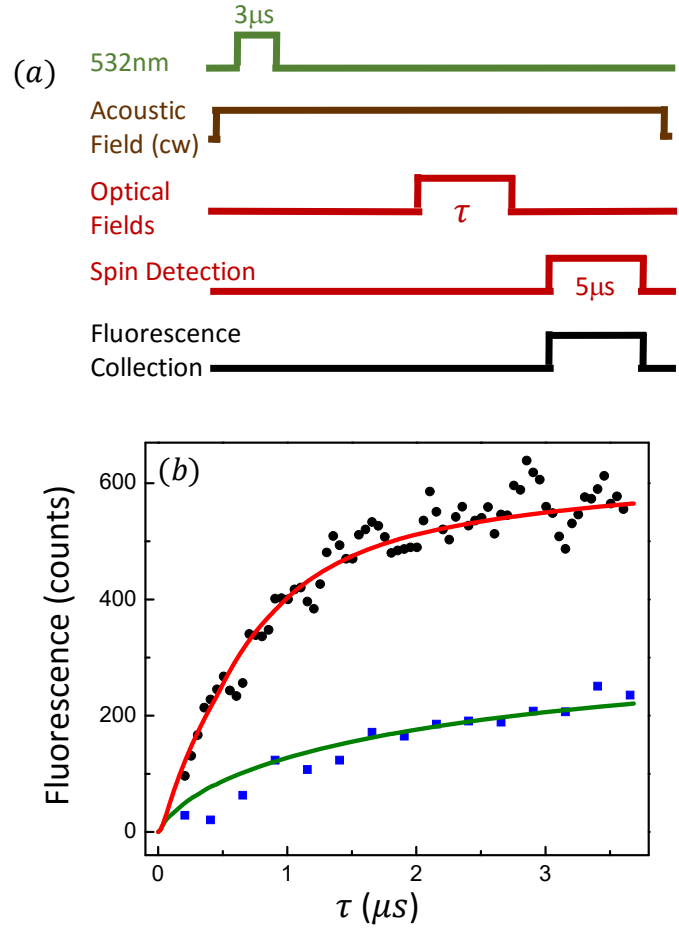

Fig. 6 (a) Pulse sequence used for the transient sideband spin transition experiment. (b) Fluorescence from state $A_2$ as a function of the optical pulse duration. Solid circles: $\omega_0+\omega_m-\omega_\pm$ satisfies the Raman resonant condition for the sideband spin transition. Solid squares: $\omega_0+\omega_m-\omega_\pm$ is 6.5 MHz detuned from the Raman resonant condition (the data are smoothed). The solid lines are the theoretical calculations discussed in the text.